\documentclass[intlimits,twoside,a4paper]{article}
\usepackage[cp1251]{inputenc}
\usepackage[eqsecnum]{cmpj3}
\usepackage{bm}
\usepackage{soul,xcolor}
\setstcolor{red}
\DeclareMathOperator\arctanh{arctanh}

\issue{2023}{26}{1}{13504}
\doinumber{10.5488/CMP.26.13504} 

\title[Impact of the {\it p}-cubic Dresselhaus term  on]
{Impact of the {\it p}-cubic Dresselhaus term on the spin Hall effect\thanks{This work is specially dedicated to professor Bertrand Berche in occasion of  his 60th birthday}}
\author[E. Santana-Su\'arez, F. Mireles]
{E. Santana-Su\'arez\refaddr{label1},
        F. Mireles\orcid{0000-0003-0506-0793}\refaddr{label2}\thanks{Corresponding author: \email{fmireles@ens.cnyn.unam.mx}}
        }
\addresses{
\addr{label1}Instituto de F\'isica, Universidad Nacional Aut\'onoma de M\'exico (UNAM), Apartado Postal 20-364, 01000 Ciudad de M\'exico, M\'exico
\addr{label2} Departamento de F\'isica, Centro de Nanociencias y Nanotecnolog\'ia, Universidad Nacional Aut\'onoma de M\'exico (UNAM),
Apartado Postal 14, 22800 Ensenada, Baja California, Mexico}

\Keywords{spin-orbit, 2DEGs, spin Hall effect, spin conductivity, spin transport 
}

\date{Received November 24, 2022}

\begin{document}
\maketitle

\begin{abstract}
It is well known that the Dresselhaus spin-orbit coupling (SOC) in semiconductor two dimensional electron gases (2DEGs)  possesses both linear and cubic in momentum contributions.\
Nevertheless, the latter is usually neglected in most theoretical studies. However, recent Kerr\
rotation experiments have revealed a significant enhancement of the cubic Dresselhaus interaction by increasing the\
drift velocities in 2DEGs hosted in GaAs quantum wells.\
Here, we present a study of the optical spin Hall conductivity in 2DEGs under the simultaneous presence of Rashba and (linear plus cubic) Dresselhaus SOC. The work was done within the Kubo formalism in linear response. We show that the coexistence of the Rashba and cubic  Dresselhaus SOC in 2DEGs promotes a strong anisotropy  of the band spin splitting which in turn leads to a very characteristic frequency dependence of the spin Hall conductivity. We find that the spin Hall conductivity response could be very sensible to   sizeable cubic-Dresselhaus coupling strength. This may be  of relevance for the optical control of spin currents in 2DEGs with non-negligible cubic-Dresselhaus SOC.

\printkeywords

\end{abstract}

\section{Introduction}

The spin-orbit coupling (SOC) is the origin of a wide range of fascinating phenomena in solid state physics \cite{Manchon}. It promotes the splitting of the conduction and valence bands in semiconductors, is the main source of spin scattering processes through the  Dyakonov--Perel~\cite{DyakanovPerel} and Elliot--Yafet~\cite{ElliotYafet1,ElliotYafet2} spin relaxation mechanisms, induces the  spin-to-charge current interconversion \cite{Li}, gives rise to the appearance of topological edge states~\cite{Hasan,Moore,Qi,Fu,Pesin} and persistent spin helix behavior~\cite{Bernevig,Koralek}, and it is a key element for the generation of the spin Hall effect~\cite{Hirsch,Zhang,Wunderlich,sinova2015spin}, among other phenomena. In two-dimensional semiconducor electron gases (2DEGs), the SOC produces a zero-field band spin splitting owing to the breaking of structural and bulk inversion symmetry, widely known as the Rashba~\cite{Rashba,Bychkov} and Dresselhaus~\cite{Dresselhaus} SOC effects. This type of SOC introduces a spin-momentum locking of the conduction electrons leading to effective magnetic fields which in turn yields to the precession of the spins.  While the Rashba Hamiltonian is linear in momentum by crystal symmetry, the Dresselhaus term possesses in general both $\bm p$-linear and $\bm p$-cubic terms contributions. Nevertheless, widespread spin-transport studies have focused on the dominant linear dependence of the Dresselhaus SOC in typical 2DEGs. Interestingly, in the $\bm p$-linear regime, and as long as the values of the Rashba and Dresselhaus coupling strengths are set equal~\cite{Schliemann}, a strong suppression of the spin relaxation can be attained, leading to the realization of the persistent spin helix phenomenon. It has been also established that the precise control of the spin state, along with the relative directions of the Rashba and both linear and non-linear Dresselhaus SOC fields are crucial for stability of a persistent spin helix state \cite{Walser, Iizasa,Zhou}.

Generally, the $\bm p$-cubic SOC of the Dresselhaus term in low-dimensional structures is neglected or assumed to be just a weak disturbance to the spin-transport properties~\cite{SpinTransport}. However, very recent Kerr rotation spectroscopy experiments on the impact of an in-plane electric field on the drifting spins in GaAs quantum wells revealed a significant enhancement of the $\bm p$-cubic Dresselhaus interaction by increasing the drift
velocities~\cite{Kunihashi}. Clearly, the $\bm p$-cubic fields introduce oscillations of the spin polarization  during the spin-transport and may cause, in conjunction with disorder, an additional spin dephasing, which are detrimental for instance to the robustness of realistic spin-transistors\cite{Datta,Chuang}. Moreover, the recent theoretical and experimental studies have put forward a significant role of the anisotropy on the spin dynamics in semiconductor narrow wires which are induced by the interaction between the linear and  cubic-Deresselhaus spin-orbit fields and
the Larmor precession under an in-plane magnetic field~\cite{Sonehara,Kawahala}. Very recent theoretical studies have emphasized a complex picture that emerges once we consider the dominant cubic-Deresselhaus SOC by exploring the superconducting correlations, the support of Majorana bound states relevant
for superconducting spintronics~\cite{Alidoust}. 
The effect of the $\bm p$-cubic term of the Dresselhaus and its interplay with the $\bm p$-linear Rashba term on the spin Hall conductivity has not been explored so far. Since the spin Hall effect is tied to the spin-charge current interconversion mediated by the Rashba and Dresselhaus SOC, there is evidently a need for a better understanding of the significance of the $\bm p$-cubic Dresselhaus contribution to the spin-transport properties.

In this paper, we use the Kubo formulation of quantum transport to study the optical spin Hall conductivity response of two-dimensional electron gases with joined Rashba and  Dresselhaus SOC effects and evaluate the impact of the cubic-Dresselhaus contribution.

\section{Spin Hall effect and SOC Hamiltonian models in 2DEGs}

The generation and manipulation of spin currents through the spin Hall effect (SHE) is of fundamental importance in spintronics. The SHE describes the flow of a spin current in a sample material in perpendicular direction to a driven longitudinal electric field in the presence of a SOC-induced effective magnetic field. Consequently, the driven electric current is deflected to the sample edges depending upon its spin orientation, causing an accumulation of spin-polarization without generating a Hall voltage  \cite{sinova2015spin}. 
The SHE can be generated by the extrinsic and intrinsic SOC mechanisms. The extrinsic mechanism is launched by impurity scattering processes in periodic crystals, which can be generated by both skew scattering \cite{Smit} and side jump \cite{Berger} processes. On the other hand, the intrinsic mechanism was first described by anomalous Hall effect (AHE) in 1954 by Karplus and Luttinger~\cite{Karplus}, but it was first related to the SHE in 2003 by Murakami, Nagaosa and Zhang \cite{Murakami,Sinova2004}. This mechanism depends on the band structure of the crystal material. The Rashba and Dresselhaus interactions are two of the main intrinsic SOC sources of the spin Hall conductivity in 2DEGs. The intrinsic SHE was first observed by Kato et al.~\cite{Kato} and Wunderlich et al. \cite{Wunderlich}.  

Since in this work we are interested in the study of spin Hall conductivity for low-dimensional semiconductor heterostructures with SOC of the Rashba and Dresselhaus type  with an emphasis on the interplay of the effect of the cubic term in the Dresselhaus SOC, herein below we discuss the main characteristics of its model Hamiltonians.  

\subsection{Rashba spin orbit coupling}
The Rashba SOC arises from the structure inversion asymmetry (SIA) in the confinement potential~$V(\mathbf{r})$ of the quantum semiconductor heterostructure that hosts the 2DEG. It is the asymmetry in  this  potential that generates an electric field $\mathbf{E}=-\nabla V(\mathbf{r})$, that in  turn  induces the coupling of the electron spins
with its momentum. The Rashba SOC is described by the Hamiltonian~\cite{Rashba,Bychkov,Bercioux}, 
\begin{equation}
    {\hat{H}}_R=\frac{\alpha}{\hbar}\left(\bm{\sigma}\times\hat{\mathbf{p}}\right)\cdot \hat{\mathbf{z}}=\frac{\alpha}{\hbar}\left({\hat{\sigma}}_x {\hat{p}}_y -{\hat{\sigma}}_y{\hat{p}}_x \right),
    \label{hamiltoniano_rashba}
\end{equation}
in which $\mathbf{\hat{p}}$ is the electron momentum in the plane of the 2DEG, ${\bm{\sigma}}$ is the vector of the Pauli matrices and $\alpha$ is the Rashba coefficient which is gate
controllable and depends on the material and on the effective electric field at the interface of the 2DEG~\cite{Bercioux}. Under this interaction and in the absence of the Dresselhaus SOC, the total Hamiltonian reads,
\begin{equation}
    {\hat{H}}={\hat{H}}_0+{\hat{H}}_R=\frac{{\hat{p}}^2}{2m^*}+\frac{\alpha}{\hbar}\left({\hat{\sigma}}_x {\hat{p}}_y -{\hat{\sigma}}_y{\hat{p}}_x \right),
    \label{hamtotal_rashba}
\end{equation}
where $\hat{H}_0$ is the free particle Hamiltonian and $m^*$ is the electron effective mass. This Hamiltonian leads to the well known spin-splitting of the electron bands in momentum space,\
\begin{equation}
    \mathcal{E}_{\pm}({\mathbf{k}})=\frac{\hbar^2k^2}{2m^*}\pm\alpha k
    \label{eigenvalor_rashba},
\end{equation}
which are characterized by the eigenvectors  
\begin{eqnarray}
\left.|{\mathbf k},\pm \right\rangle  =\frac{\rm{e}^{i{\mathbf{k}}\cdot\mathbf{r}}}{\sqrt{2}}\left(\begin{array}{cc}
      1   \\
      \mp \ri{\rm{e}}^{\ri\theta}  
    \end{array}\right), \quad \text{with} \quad
    \theta=\arctan\left(\frac{k_y}{k_x}\right).
    \label{eigenvector_rashba} 
\end{eqnarray}

\noindent where ${\mathbf{k}}=(k_x,k_y)$ and $k=|{\mathbf{k}}|$.

\subsection{Dresselhaus spin orbit coupling}
The three-dimensional SOC correction for a free particle Hamiltonian is due to the bulk inversion asymmetry (BIA) observed in semiconductors with zinc-blende cristalline structure, and it is given by the Dresselhaus Hamiltonian \cite{Dresselhaus}
\
\begin{equation}
    {\hat H}_{D}^{3D}=\gamma \left[{\hat p}_x {\big(\hat{p}_y ^2-p_z^2\big){\hat \sigma }_x+{\hat p}_y \big(p_z^2-{\hat p }_x ^2\big){\hat \sigma}}_y+\hat{p}_z \big({\hat p }_x ^2-{\hat p }_y ^2\big){\hat \sigma }_z\right],\
    \label{hamiltoniano_dresselhaus3d}
\end{equation}
where the $\gamma$ coefficient is a material dependent constant and gives the strength of the Dresselhaus SOC. For a quantum well potential profile along the $z$-direction, a two-dimensional Hamiltonian ${\hat H}_D^{2D}$ is obtained after taking the expectation value with the wave function of the quantum well ground state along such a direction,  $\langle \hat{H}_D^{3D}\rangle$, which after the use of  $\langle{p_z}\rangle= 0$ due to the quantum well symmetry, in conjunction with  $\langle{p_z}^2/\hbar^2\rangle\sim (\piup/L_z)^2$, with $L_z$ being the width of the quantum well layer, the Dresselhaus SOC Hamiltonian for the 2DEG reduces to \
\
\
\begin{equation}
    \hat{H}_{D}^{2D}=\hat{H}_D+\hat{H}_{D^3}=\frac{\beta}{\hbar}\big({\hat{p}}_y {\hat{\sigma}}_y-{\hat{p}}_x {\hat{\sigma}}_x\big)+\frac{\gamma}{\hbar^3}\big({\hat{p}}_x  {\hat{p}}_y ^2{\hat{\sigma}}_x-{\hat{p}}_y  {\hat{p}}_x ^2{\hat{\sigma}}_y\big),\
    \label{hamiltoniano_dresselhaus3d2}
\end{equation}
where $\beta=\gamma \langle{p_z}^2/\hbar^2\rangle$. Hence, in general the Dresselhaus SOC in 2DEGs has both a linear ${\hat H}_D$ and a cubic ${\hat H}_{D^3}$ contribution. The latter is usually neglected with the argument that $\langle{p_z^2}\rangle \gg {p}_x , {p}_y$.

\subsection{Joined Rashba and Dresselhaus SOC}

As we mentioned, here we are interested in the effect of the cubic term  of Dresselhaus SOC in 2DEG's under the joined action of the Rashba and Dresselhaus SOC. The total Hamiltonian of this system can be written as\
\begin{equation}
    \hat{H}=\hat{H}_0+\hat{H}_R+\hat{H}_D^{2D}=\frac{{\hat{p}}^2}{2m^*}+\frac{\alpha}{\hbar}\big({\hat{\sigma}}_x {\hat{p}}_y -{\hat{\sigma}}_y{\hat{p}}_x \big)+\frac{\beta}{\hbar}\big({\hat{p}}_y {\hat{\sigma}}_y-{\hat{p}}_x {\hat{\sigma}}_x\big)+\frac{\gamma}{\hbar^3}\big({\hat{p}}_x  {\hat{p}}_y ^2{\hat{\sigma}}_x-{\hat{p}}_y  {\hat{p}}_x ^2{\hat{\sigma}}_y\big).\
    \label{hamiltoniano_total}
\end{equation}
This Hamiltonian leads to the energy band dispersion,\
\begin{equation}
\mathcal{E}_{\nu}(\mathbf{k},\, \theta)=\frac{\hbar^2k^2}{2m^*}+\nu\Delta(k, \theta)k,\
    \label{eigenvalores}
\end{equation}
in which $\nu=\pm 1$, $\theta$ is given in equation (\ref{eigenvector_rashba}), and\
\begin{equation}
    \Delta(k, \theta)=\sqrt{\alpha^2+\beta^2-2\alpha\beta \sin2\theta+\gamma k^2 \sin2\theta\left(-\alpha +\beta \sin{2\theta}+\frac{\gamma}{4}k^2 \sin2\theta\right)}.\
    \label{delta_k_theta}
\end{equation}
\
The eigenstates for the Hamiltonian (\ref{hamiltoniano_total}) are given by\
\begin{eqnarray}
   |{\mathbf k},\nu\rangle=\frac{{\rm e}^{\ri{\mathbf{k}} \cdot\mathbf{r} }}{\sqrt{2}}\left(\begin{array}{cc}
     1 \\  
      \frac{\ri \nu}{k\Delta(k, \theta)}(\alpha k_+ -\ri\beta k_--\gamma k_+ k^2 \sin{\theta} \cos\theta) 
      \label{eigenvectores}
   \end{array}\right),
\end{eqnarray}
with $k_{\pm}=k_x\pm \ri k_y$.

\section{Optical spin Hall conductivity (SHE)}

Consider now a 2DEG with joined Rashba and Dresselhaus SOC as in (\ref{hamiltoniano_total}) and let us seek for  the response of applying an uniform electric field of frequency $\omega$ and aligned along the ${\hat y}$ direction. If the perturbation starts at time $t_0=0$, then the optical spin Hall conductivity of spins $z$-oriented electrons flowing along the ${\hat x}$ direction can be described using the Kubo formalism of quantum transport. It reads explicitly,  
\begin{equation}
    \sigma^{s_z}_{xy}(\omega)=\frac{e}{\hbar A (\omega+\ri\eta)}\int_{0}^{\infty}\rd t\, {\mathrm{e}}^{\ri(\omega+\ri\eta)t}\sum_{\bm{k}, \nu} f({\cal{E}}_{\nu})_{T=0}{\langle\Psi_{\bm{k},\nu}(\bm{r})|[\hat{{\cal J }}^{s_z}_x(t), {\hat{v}}_y(0)]|\Psi_{\bm{k}, \nu}(\mathbf{r})\rangle},\
    \label{kubo_espin}
\end{equation}
\noindent where $f({\cal{E}}_{\nu})_{T=0}$ is the Fermi-Dirac distribution function evaluated at zero temperature,  ${\hat{\mathbf{\cal{J}}}}^{s_z}_x(t)$ is the conventional spin-current operator written in the interaction representation, ${\hat v}_y(0)$ is the velocity operator at $t_0 =0$, $|\Psi_{\bm{k},\nu}(\bm{r})\rangle$ are the eigenvectors of the non-perturbed Hamiltonian, $e$ is the charge of the electrons, $A$ is the area of the sample, and $\eta$  shows that the perturbation is turned on adiabatically. This ensures a causal response, but also can be interpreted as a  phenomenological parameter that  characterizes the disipation effects due to the scattering by disorder. In analogy with the charge current definition, the conventional spin-current operator  written in the Schr\"odinger representation is expressed through the anti-commutator, 
\begin{equation}
{\hat{\mathbf{\cal{J}}}}^{s_z}_x=\frac{1}{2}\{{\hat{v}}_x,{\hat{s}}_z\},
\label{densidad_corriente_espin}
\end{equation}
with ${\hat{s}}_z=\frac{\hbar}{2}{\hat{\sigma}}_z$ and the $x$-component of electron velocity operator is ${\hat{v}}_x(0)=\frac{\partial \hat{H}}{\partial {\hat{p}}_x}
    \label{velocidad_x}$. \

\subsection{Pure Rashba coupling}

\begin{figure}
	\centering
         \includegraphics[width=7.2cm]{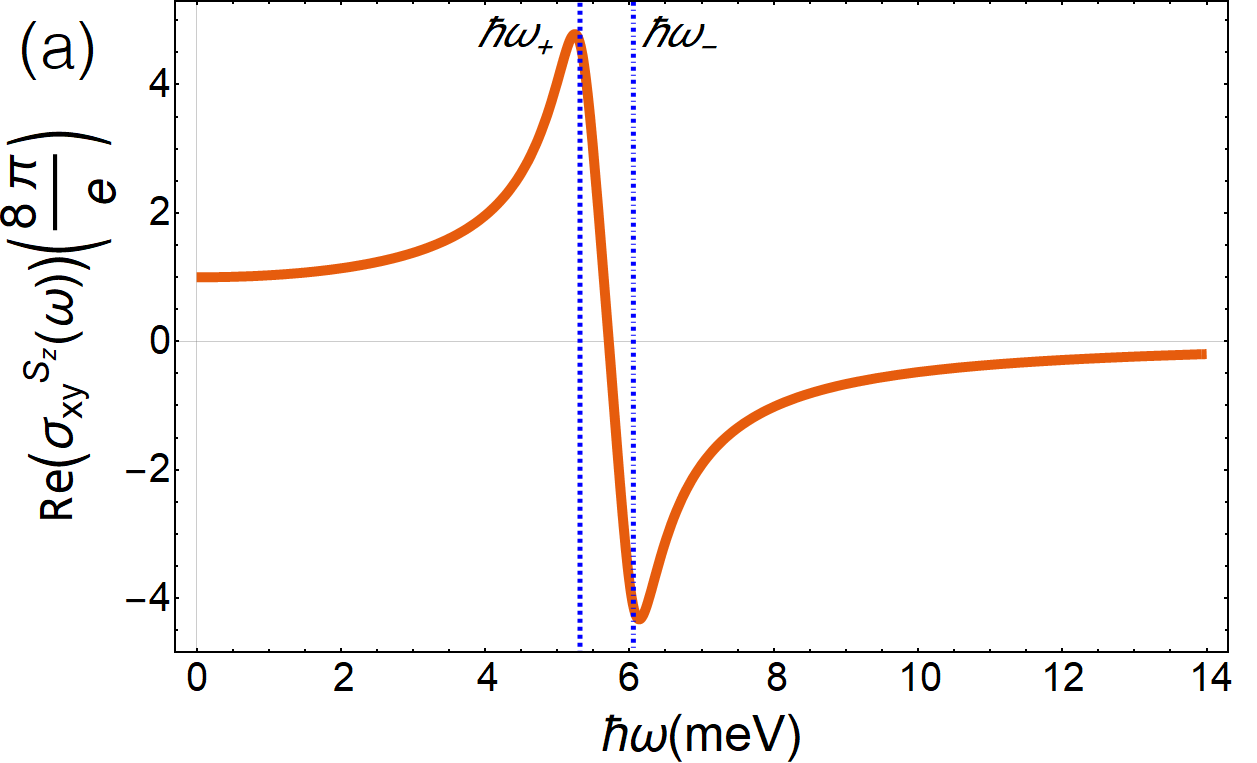}  \hspace{0.2cm}
         \includegraphics[width=7.2cm]{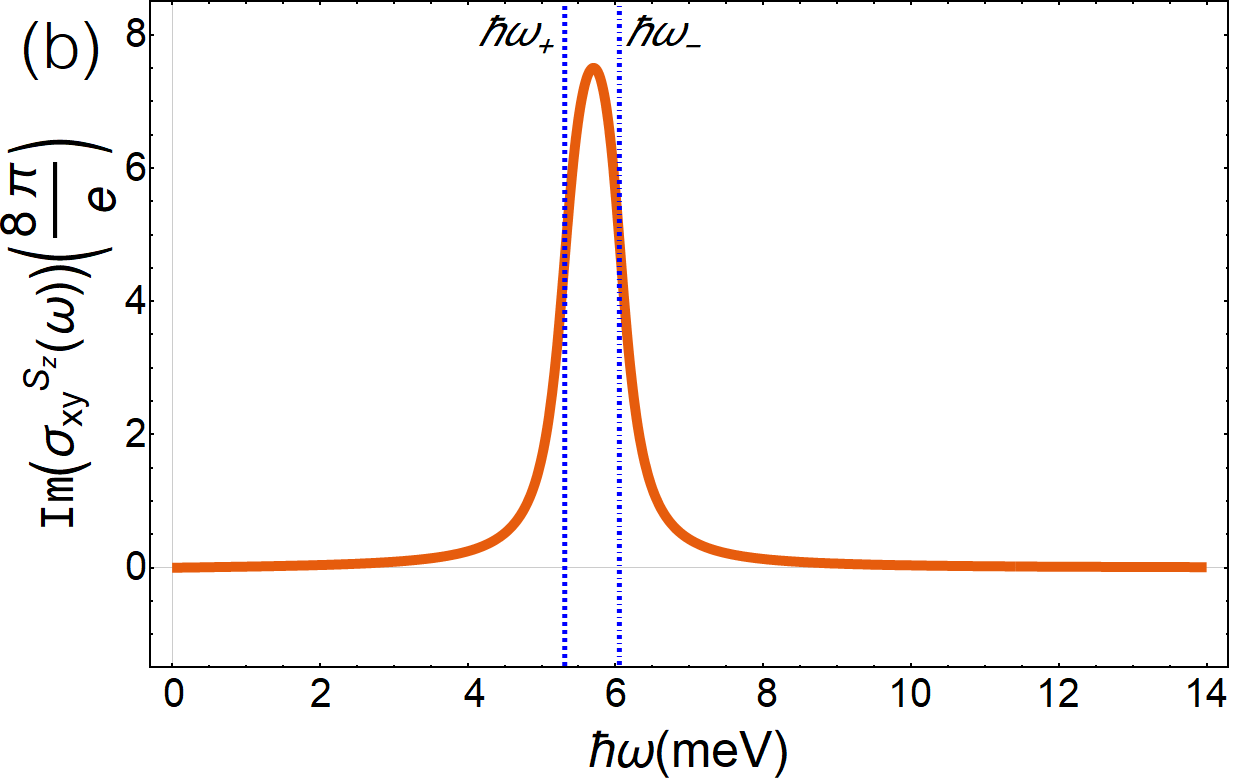}
         \label{fig:rsoc_im}
        \caption{(Colour online) Spin Hall conductivity for an InAs-based 2DEG under pure Rashba SOC. (a)~Real part, and (b) imaginary contribution. See text for the description of the characteristic energies~$\hbar \omega_{\pm}$.}
        \label{fig:rsoc_conductivity}
\end{figure}

It is illustrative to revisit the result obtained for the frequency dependent spin Hall conductivity in the absence of Dresselhaus interaction, considering the Rashba SOC only. For this situation, the expectation value of the ``spin current-charge current'' correlation function leads simply to\
\begin{equation}
    \langle{\mathbf{k}, \nu|[{\hat{\mathbf{\cal{J}}}}^{s_z}_x(t), {\hat{v}}_y(0)]|\mathbf{k}, \nu}\rangle=\frac{\ri\nu \alpha\hbar}{m^*k}k_x^2\,\cos\left(\frac{2\alpha k t}{\hbar}\right), 
    \label{expectacion_rashba}
\end{equation}
from which, after making the substitution 
$$\sum_{\mathbf{k}, \nu}\nu f(\mathcal{E}_{\nu})_{T=0}  \rightarrow -\int_{0}^{2\piup} \int_{k_{\rm F_+}}^{k_{\rm F_-}}\frac{Ak}{(2\piup)^2}\rd k\,\rd\theta,$$
and carrying out the time-integration in the Kubo formula, leads to the frequency dependent spin Hall conductivity, 
\begin{equation}
    \sigma^{s_z}_{{xy}}(\omega)=\frac{e}{8\piup}-\frac{\hbar^3 e}{32\piup \alpha^2 m^*}(\omega+\ri\eta)\,\left\lbrace \arctanh\left[\frac{2\alpha k_{\rm F_-}}{\hbar (\omega+\ri\eta)}\right]-\arctanh\left[\frac{2\alpha k_{\rm F_+}}{\hbar (\omega+\ri\eta)}\right]\right\rbrace,\
    \label{conductividad_frec_rashba}
\end{equation}
where $k_{{\rm F}_{\pm}}$ are the (Fermi) wave numbers for the spin-splitted $\nu=\pm$ bands at the Fermi energy $\mathcal{E}_{\rm F}$, defined by
\begin{equation}
    k_{{\rm F}_{\pm}}=\mp \frac{\alpha m^*}{{\hbar}^2}+\sqrt{\frac{{\alpha}^2 {m^*}^2}{{\hbar}^4}+\frac{2 m^*}{{\hbar}^2}\mathcal{E}_{\rm F}},
    \label{kf_rashba}
\end{equation}
with $\mathcal{E}_{\rm F}={{\hbar}^2 {k_{\rm F}}^2}/{2m^*}$ and $k_{\rm F}=\sqrt{2 \piup n_e}$, being $n_e$ the electron density of the 2DEG. Interestingly,  in the static limit, $\omega\rightarrow 0$ and when the disorder is negligible $\eta\rightarrow 0$, the spin Hall conductivity does not depend on the material parameters, and it is given by
\begin{equation}
    \sigma^{s_z}_{{xy}}(0)=\frac{e}{8\piup}.\
    \label{conductividad_0_rashba}
    \end{equation}
However, it is known that this contribution is eventually cancelled by short-range disorder scattering\
because the driven spin-current is proportional to the dynamics of the spins, which should necessarily vanish in the steady state regime~\cite{sinova2015spin}. Therefore, the constant result (\ref{conductividad_0_rashba})    
is valid in the ideal system without electron-electron interactions and without the scattering with impurities or disorder. In figure~\ref{fig:rsoc_conductivity} it is shown a frequency dependence of the real part (left-hand panel) and imaginary part (right-hand panel) of the spin Hall conductivity for a typical 2DEG in InAs quantum well under Rashba SOC. 

\subsection{Joined Rashba and linear Dresselhaus coupling}

Before we study the impact of the cubic Dresselhaus term on the spin Hall conductivity it is useful first to discuss the coexistence of the Rashba and linear Dresselhaus interactions. Starting with the   Hamiltonian  (\ref{hamiltoniano_total}) with the cubic Dresselhaus parameter $\gamma=0$, one arrives at the ``spin current-charge current'' correlation function, 
\begin{equation}
\langle{\mathbf{k}, \nu|[{\hat{\mathbf{\cal{J}}}}^{s_z}_x(t), {\hat{v}}_y(0)]|\mathbf{k}, \nu}\rangle=\frac{- \ri \nu\hbar k_x^2}{m^*k\Delta_0(\theta)}\big(\alpha^2-\beta^2\big)\cos\left[\frac{2k\Delta_0(\theta) t}{\hbar}\right],
\end{equation}
with $\Delta_0(\theta)\equiv\Delta(0,\theta)=\sqrt{\alpha^2+\beta^2-2\alpha\beta\sin{2\theta}}$, which is contrary to the pure Rashba case, it is angular dependent due to the interplay between the Rashba and linear Dresselhaus couplings. This fact together with the angular anisotropy of its energy  band dispersion,  ${\mathcal E}_{\nu}({\mathbf k},\, \theta)=({\hbar^{2} k^2}/{2m^*})+\nu\Delta_0(\theta)k,$  entails an anisotropic spin-splitting, ${\mathcal E}_{+}-{\mathcal E}_{-}=\Delta_{0}(\theta)k$, as well as an angular dependence of Fermi wave numbers for the spin-splitted bands $k_{{\rm F}_{\pm}}\rightarrow k_{{\rm F}_{\pm}}(\theta)$. All together avoid us to arrive at an analytical expression for the spin Hall conductivity, and get instead,
\begin{equation}
    \sigma^{s_z}_{{xy} }(\omega)=\frac{-e\big(\beta^2-\alpha^2\big)}{m^*(2\piup)^2}\int_{0}^{2\piup} \int_{k_{\rm F_{+}}(\theta)}^{k_{\rm F_{-}}(\theta)}\frac{k^2 \cos^2\theta}{\Delta_0(\theta)}\frac{1}{(\omega+\ri\eta)^2-\left[{2k\Delta_0(\theta)}/{\hbar}\right]^2}\, \rd k\,\rd \theta,
    \label{integral_conductividad_rdlineal}
\end{equation}
that should be integrated numerically in $\theta$ after using $k_{\rm F_-}-k_{\rm F_+}={2m^*\Delta_0(\theta)}/{\hbar^2}$.  Now, if we take the static limit ($\omega\rightarrow 0$) for a 2DEG in the clean limit ($\eta\rightarrow 0$), together with the reasonable assumption that ${2k\Delta_0(\theta)}/{{\cal E}_{F}}\ll 1$, the spin Hall conductivity  (\ref{integral_conductividad_rdlineal}) reduces to
\begin{equation}
    \sigma^{s_z}_{xy}(0)=
    \frac{\big(\beta^2-\alpha^2\big)e}{8\piup^2}\int_{0}^{2\piup}\frac{\cos^2\theta}{\Delta^2(\theta)}\,\rd \theta,\
\end{equation}
that once we integrate over $\theta$, the spin Hall conductivity in the static limit yields\cite{Sinitsyn,Shen},
\begin{equation}
    \sigma^{s_z}_{xy}(0)=\frac{e}{8\piup}\frac{\alpha^2-\beta^2}{|\alpha^2-\beta^2|}.
    \label{conductividad_rdlineal_estatico}
\end{equation}

\begin{figure}
	\centering
 \hspace{-0.5cm}
    \includegraphics[width=7.3cm]{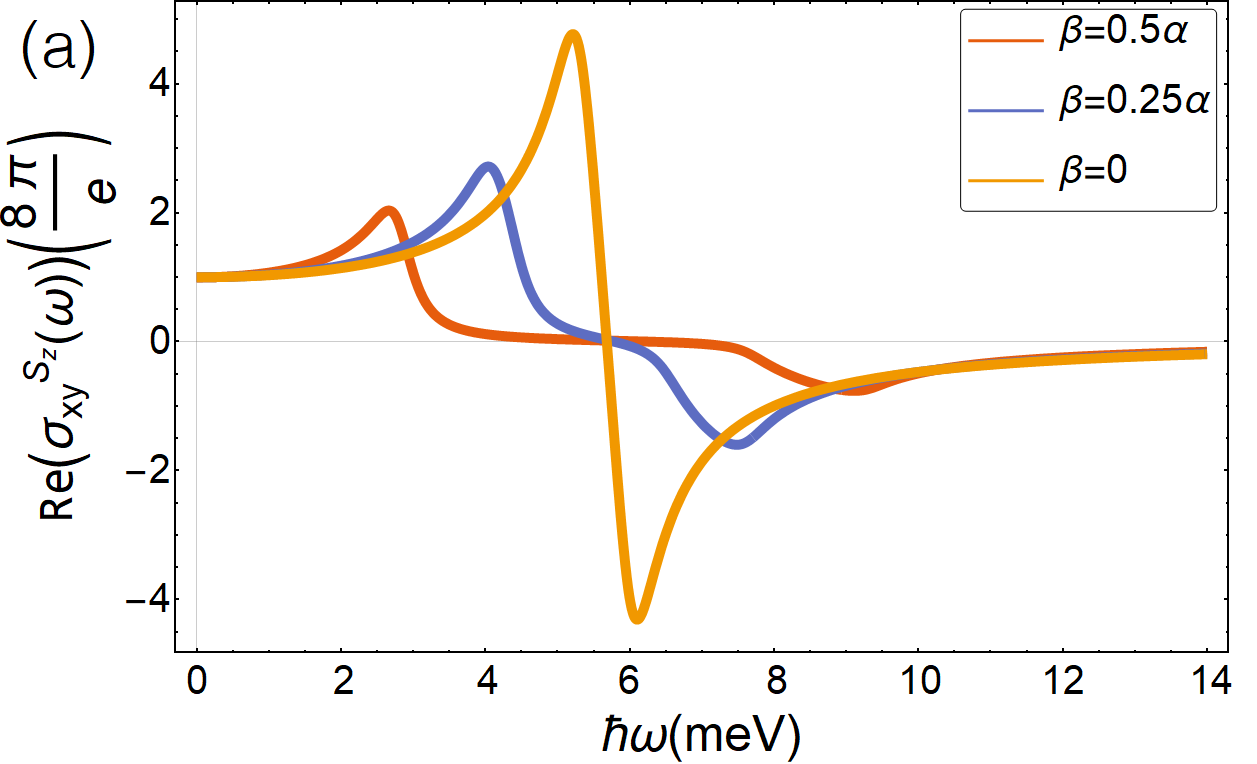}  
    \includegraphics[width=7.3cm]{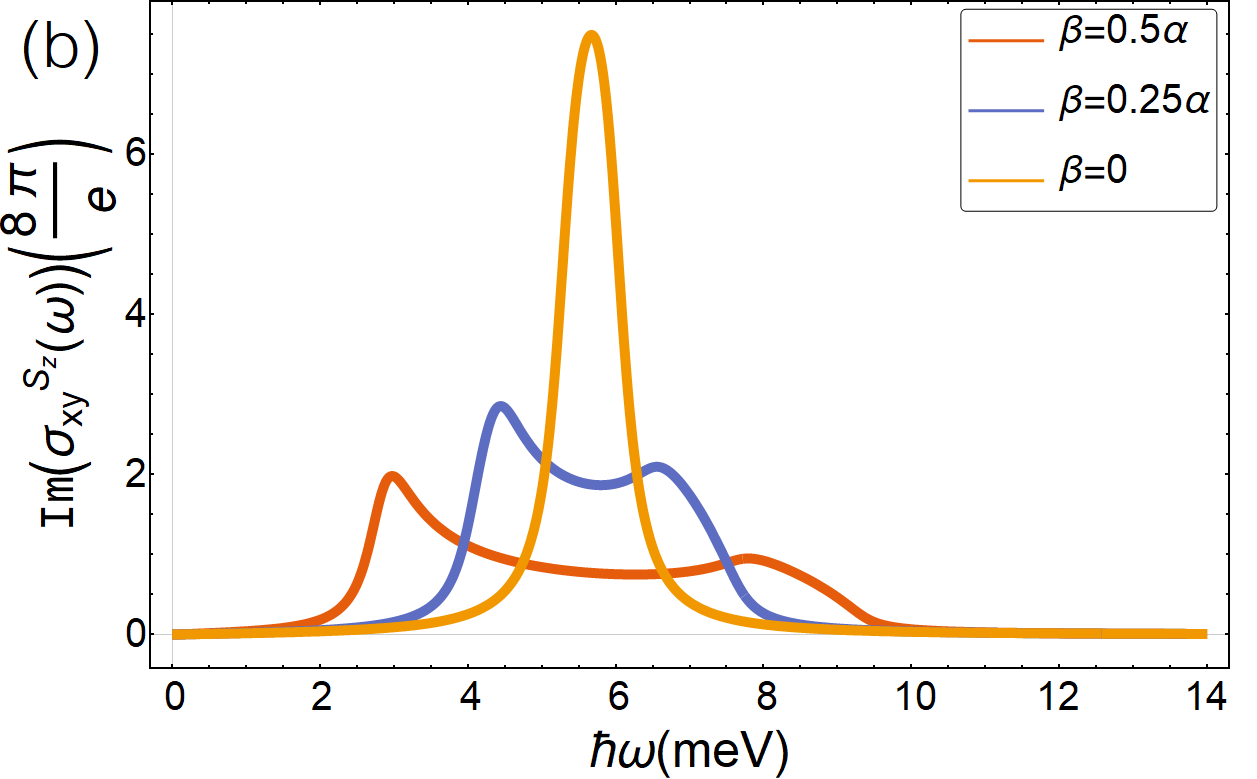}
    \caption{(Colour online) Spectra for (a) the real and (b) the imaginary part of the spin Hall conductivity in a InAs-based 2DEG under joined Rashba and lineal Dresselhaus SOC. The plots were calculated using the values  $\alpha=1$.$6\times{10}^{-11}$ eV m, electron density  $n_e=5\times{10^{15}}$~m$^{-2}$ and $\hbar\eta=0$.$25\times{10^{-3}}$ meV. 
    Three cases are shown for different $\beta$ values:  $\beta=0.5\alpha$ (orange),  $\beta=0.25\alpha$ (blue) and $\beta=0$ (yellow).}
    \label{fig:conductividad_rdlineal}
\end{figure}

The real and imaginary parts of the optical spin Hall conductivity given by equation (\ref{integral_conductividad_rdlineal}) are shown in figure~\ref{fig:conductividad_rdlineal}. For both plots of the spectral response, it can be seen that if the parameter $\beta$ increases, the spectrum widens and the maximum and minimum of the spin conductivity tends to separate from each other. For the real part (left-hand panel) the spin Hall conductivity in the static limit approaches to the constant $e/8\piup$ value, whiles for higher frequencies, the conductivity approaches zero. By constrast, the imaginary part (right-hand panel), for both static limit and large frequencies, the conductivity drops to zero.

\begin{figure}
    \includegraphics[width=7.5cm,height=4.5cm]{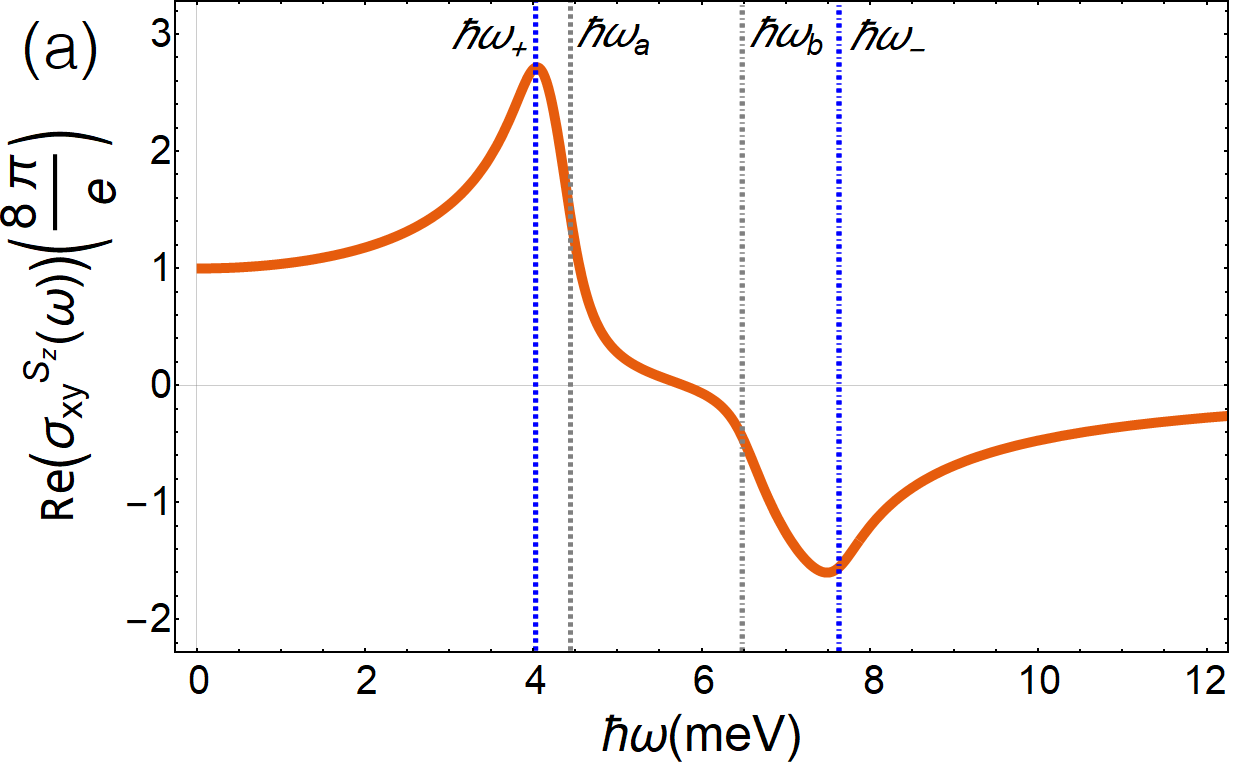}  
    \includegraphics[width=7.5cm,height=4.5cm]{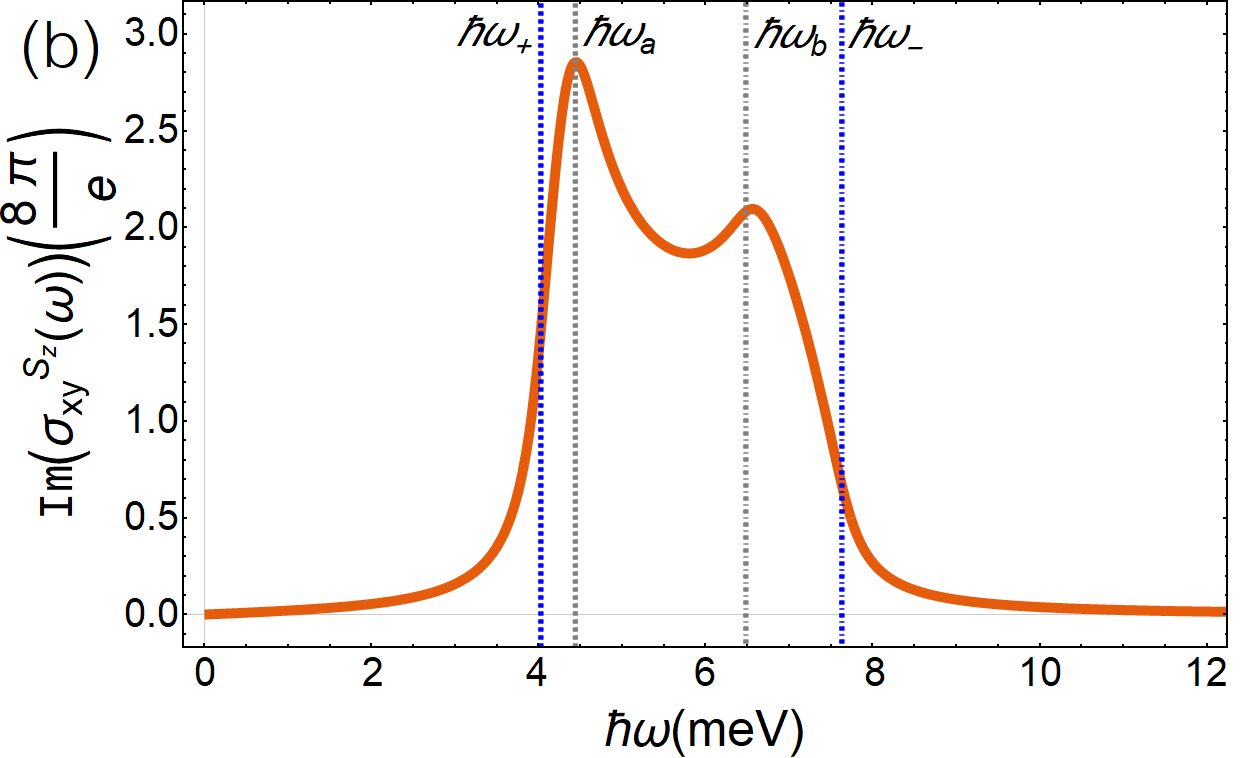}
    \hspace{-0.5cm}
    \includegraphics[width=7.5cm,height=4.5cm]{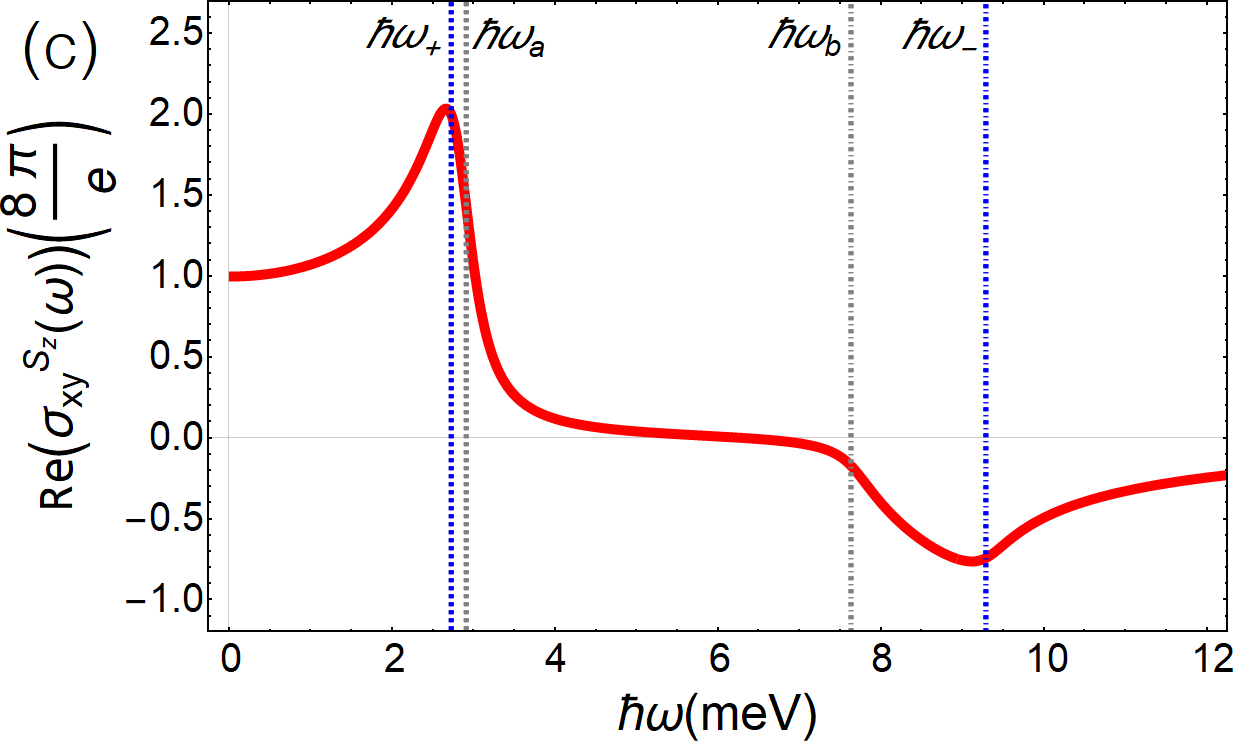}
    \includegraphics[width=7.5cm,height=4.5cm]{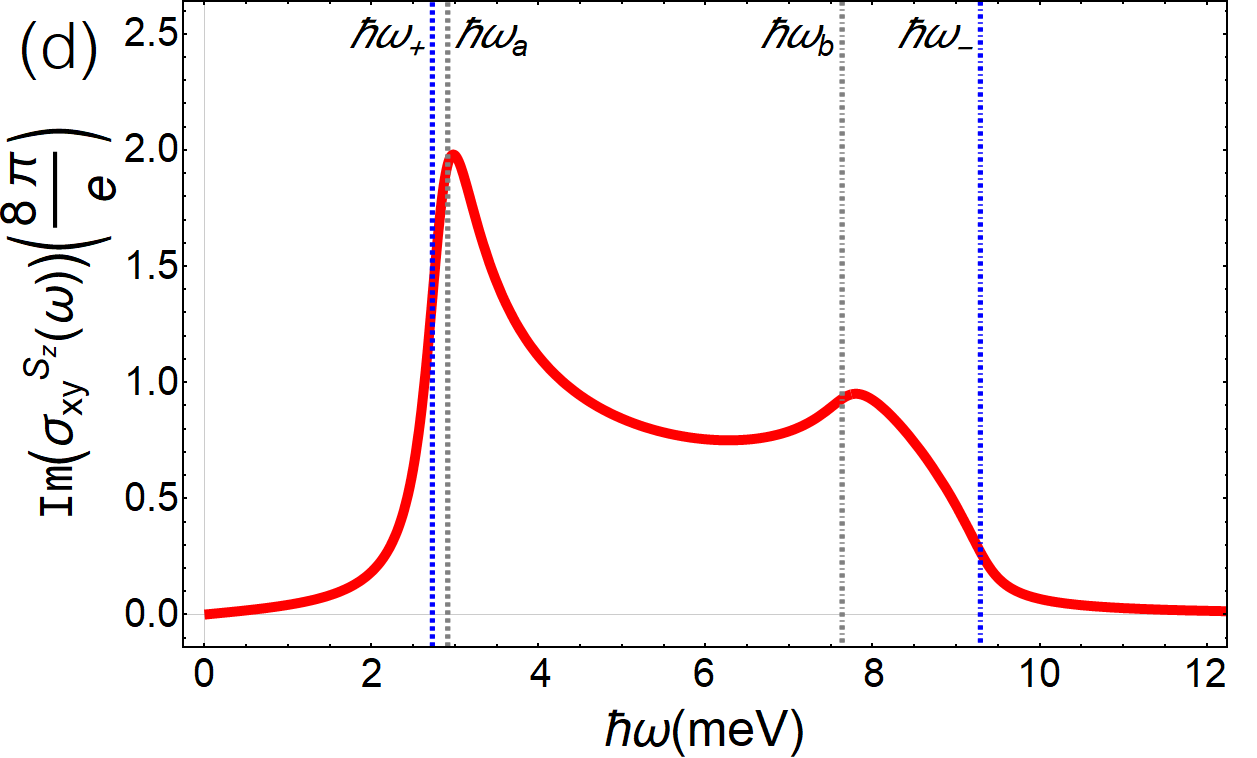}
    \caption{(Colour online) Spin Hall conductivity spectra for a system with joined Rashba and lineal Dresselhaus SOC for $\beta=0$.$25\alpha$ (a and b) and $\beta=0$.$50\alpha$ (c and d). In this spectra the dashed blue line represents the $\hbar\omega_+$ value, the dashed gray line represents $\hbar\omega_a$, the dotted gray line  is the frequency  $\hbar\omega_b$ and the dotted blue line represents the frequency $\hbar\omega_-$.}
    \label{fig:rdfrec}
\end{figure}

These properties of the optical spin conductivity respond to the  spin-split anisotropy of the bands for joined Rashba and linear Dresselhaus SOC, and can alternatively be understood through the calculation of the joint density of states (JDOS)\cite{Wong}. 
Note that the angular dependence adds four  characteristic frequencies, two related to the absorption edges in the spectrum denoted by $\omega_+$ and $\omega_-$, and two more responsible for the absorption and high-density peaks (singularities) of the JDOS at photon frequencies $\omega_a$ and $\omega_b$, respectively. These frequencies are connected to the angular energy spin-splitting of the bands given by 
\begin{eqnarray}
    \hbar\omega_+& = &  2k_{\rm F_+}\left(\frac{\piup}{4}\right)\Delta_0\left(\frac{\piup}{4}\right),\\ 
    \hbar\omega_-& = &  2k_{\rm F_-}\left(\frac{3\piup}{4}\right)\Delta_0\left(\frac{3\piup}{4}\right), \\
    \hbar\omega_a &= & 2k_{\rm F_-}\left(\frac{\piup}{4}\right)\Delta_0\left(\frac{\piup}{4}\right), \\
    \hbar\omega_b & =& 2k_{\rm F_+}\left(\frac{3\piup}{4}\right)\Delta_0\left(\frac{3\piup}{4}\right),
    \label{omegas}
\end{eqnarray}
which gives meaning to the the structure of the spectra profile of the spin Hall conductivity (see figure~\ref{fig:rdfrec}). Using the  equations above, we find that for $\beta=0.25\alpha$, the characteristic energies for such frequencies in an InAs-based 2DEG are $\hbar\omega_+= 4.05$~meV, $\hbar\omega_a=4.46$~meV, $\hbar\omega_b=6.51$~meV and $\hbar\omega_-=7.66$~meV. For $\beta=0.50\alpha$ the important frequencies are $\hbar\omega_+= 2.74$~meV, $\hbar\omega_a=2.93$~meV, $\hbar\omega_b=7.68$ meV and $\hbar\omega_-=9.34$~meV. These energies are depicted with dotted lines in figure~\ref{fig:rdfrec}. It is clear that  such frequencies basically correspond to the peaks, minuma, as well as to the changes in the slope of the optical spin Hall conductivity.  

\subsection{Rashba with linear plus cubic Dresselhaus couplings}
\label{subsec:joinedrashbadresselhaus}

In this section we analise the spin Hall conductivity response using the full 2D Hamiltonian given in the equation (\ref{hamiltoniano_total}) that takes into account the cubic Dresselhaus term ($\gamma\neq0$). The exact eigenvalues and the eigenvectors for this system are provided by equations (\ref{eigenvalores}) and (\ref{eigenvectores}). It is clear that this general model that also includes the cubic contribution of Dresselhaus SOC is much more complicated, and as a consequence, it turns out to be even more intricate to be solved for the spin Hall conductivity analytically than the previous models, and at a certain point we shall appeal to numerical computations. 

It can be shown that the ``spin current-charge current'' correlation function (see Appendix) associated to  the Hamiltonian  (\ref{hamiltoniano_total}) is given by, 
\begin{equation}
\langle{\mathbf{k}, \nu|[{\hat{\mathbf{\cal{J}}}}^{s_z}_x(t), {\hat{v}}_y(0)]|\mathbf{k}, \nu}\rangle=\frac{\nu \ri\hbar k_x^2}{m^*k\Delta(k,\theta)}\cos\left[\frac{2k\Delta(k,\theta) t}{\hbar}\right]
{\cal G}(k,\theta),
\label{correlation_function}
\end{equation}
where $\Delta(k,\theta)$ is defined in   (\ref{delta_k_theta}), and 
\begin{equation}
{\cal G}(k,\theta)=\beta^2-\alpha^2 +\gamma \left(\alpha \sin2\theta+\beta \cos2\theta\right)k^2- \frac{\gamma^2}{4}\sin(2\theta) k^4.
\end{equation}

Therefore, the Kubo formula for the joint Rashba and cubic Dresselhaus SOC reads explicitly,
\begin{equation}
    \sigma^{s_z}_{xy}(\omega)=\frac{-\ri e}{   m^*(\omega+\ri\eta)}\int_{0}^{\infty}\rd t\, {\mathrm{e}}^{\ri(\omega+\ri\eta)t}\int_{0}^{2\piup} \int_{k_{\rm F_{+}}(\theta)}^{k_{\rm F_{-}}(\theta)}\frac{k^2}{(2\piup)^2}\rd k\,\rd\theta\frac{ \cos^2\theta}{\Delta(k,\theta)}\cos\left[\frac{2k\Delta(k,\theta) t}{\hbar}\right]{\cal G}(k,\theta).\
    \label{kubo_cubicD}
\end{equation}
Even though this expression for the optical spin Hall conductivity is exact, it is not amenable for analytical integration, and therefore a numerical integration should be implemented. However,  before we proceed further in analyzing its numerical behavior, it is illustrative to consider some assumptions that will lead us to approximate expressions in the static regime and vanishing $\eta$. 

First we notice that the ratio 
\begin{equation} 
\frac{\mathcal{G}(k, \theta)}{\Delta^2(k, \theta)} \simeq  \frac{\mathcal{G}({k_{\rm F_0}(\theta)}, \theta)}{\Delta^2({k_{\rm F_0}(\theta)}, \theta)}, \quad {\text{with}  \quad k_{\rm F_0}(\theta)=-\frac{\Delta_0(\theta) m^*}{\hbar^2}+\sqrt{\frac{\Delta_0^2(\theta) {m^{*}}^2}{\hbar^4}+\frac{2 m^*}{h^2}\mathcal{E}_{\rm F}},}
\end{equation}
which allows us to rewrite ( \ref{kubo_cubicD}) as 
\begin{eqnarray}
    \sigma^{s_z}_{{xy}}(\omega)=&-&\frac{e\hbar^3}{8\piup^2 m}\int_{0}^{2\piup}\cos^2\theta \frac{\mathcal{G}({k_{\rm F_0}}, \theta)}{\Delta({k_{\rm F_0}}, \theta)}\frac{(\omega+\ri\eta)}{\left[2\Delta({k_{\rm F_0}}, \theta)\right]^3}  \,\ln\left[\frac{\hbar(\omega+\ri\eta)-2\Delta({k_{\rm F_0}}, \theta)k_{\rm F_+}}{2\Delta({k_{\rm F_0}},\theta)k_{\rm F_-}-\hbar(\omega+\ri\eta)}\right]\rd \theta \nonumber \\
    &+&\frac{e}{8\piup^2}\int_{0}^{2\piup}\cos^2\theta \frac{\mathcal{G}({k_{\rm F_0}}, \theta)}{\Delta^2({k_{\rm F_0}}, \theta)}\rd \theta, 
    \label{sigmardddt}
\end{eqnarray}
where the Fermi wave numbers are now defiend as
\begin{equation}k_{{\rm F}_{\pm}}=\mp \frac{m^* \Delta({k_{\rm F_{0}}}, \theta)}{\hbar^2}+\sqrt{\frac{{m^*}^2 \Delta{^2}(k_{\rm F_{0}}, \theta)}{\hbar^4}+\frac{2m^*}{\hbar^2}{\cal E}_{\rm F}}.\
    \label{limiterd3}
\end{equation}
Hence, in static limit ($\omega\rightarrow0$ and $\eta\rightarrow0$), the spin Hall conductivity reduces to\
\begin{equation}
    \sigma^{s_z}_{{xy}}(0)=\frac{e}{8\piup^2}\int_{0}^{2\piup}\cos^2\theta \,\, \frac{\mathcal{G}({k_{\rm F_0}}, \theta)}{\Delta^2({k_{\rm F_0}}, \theta)}\rd \theta,\
    \label{gammalimest}
\end{equation}
which can be further simplified for the case of a realtively large quantum well, that is considering $\gamma k_{\rm F}^2 \gg \beta$ and $\alpha\gg\gamma  k_{\rm F}^2$, for instance in an InAs quantum well. Under these considerations, the equation~\eqref{gammalimest} leads to
\begin{eqnarray}
    \sigma_{xy}^{s_z}(0)&=&\frac{e}{8\piup^2}\int_0^{2\piup}\rd \theta \cos^2\theta\,\,\frac{-\alpha^2+\beta^2+\alpha\gamma k_{\rm F}^2 \sin2\theta}{\alpha^2+\beta^2-2\alpha\beta \sin2\theta-\alpha\gamma k_{\rm F}^2 \sin2\theta} \nonumber\\
    &\approx & \frac{e}{8\piup}\left[\frac{\gamma k_{\rm F}^2}{2\beta+\gamma k_{\rm F}^2}+\frac{2\beta}{\big(2\beta+\gamma k_{\rm F}^2\big)}\frac{\big(\alpha^2-\beta^2\big)}{|\alpha^2-\beta^2|}\right],
\end{eqnarray}
and since $({\alpha^2-\beta^2})/{|\alpha^2-\beta^2|}=\text{sign}(\alpha-\beta)$, then we can write
\begin{equation}
    \sigma_{xy}^{s_z}(0)\approx\frac{e}{8\piup}\left\{1-\frac{2\beta}{2\beta+\gamma k_{\rm F}^2}\left[1-\text{sign}(\alpha-\beta)\right]\right\}.
\end{equation}
Note that if $\gamma=0$, this approximation for the static spin Hall conductivity  reduces, as we expect, to the constant value found for a 2DEG under joined Rashba and linear-Dresselhaus SOC, see equation~(\ref{conductividad_rdlineal_estatico}). Similarly, it reduces to $e/8\piup$, equation~(\ref{conductividad_0_rashba}), for the case of  vanishing linear-Dresselhaus term ($\beta\rightarrow 0$) and finite Rashba ($\alpha\ne0$) and finite cubic-Dresselhaus ($\gamma\ne0$) SOC terms.\
\begin{figure}[h]\
    \includegraphics[width=7.5cm]{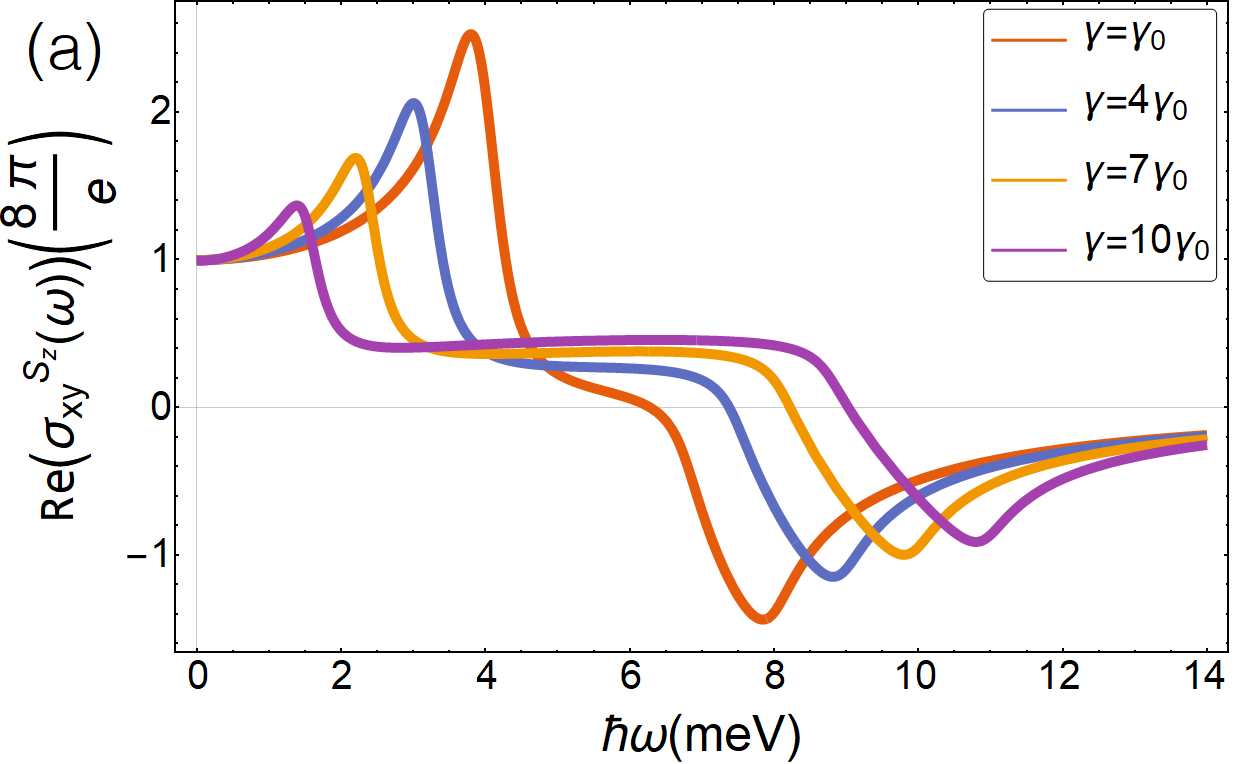}  
    \includegraphics[width=7.5cm]{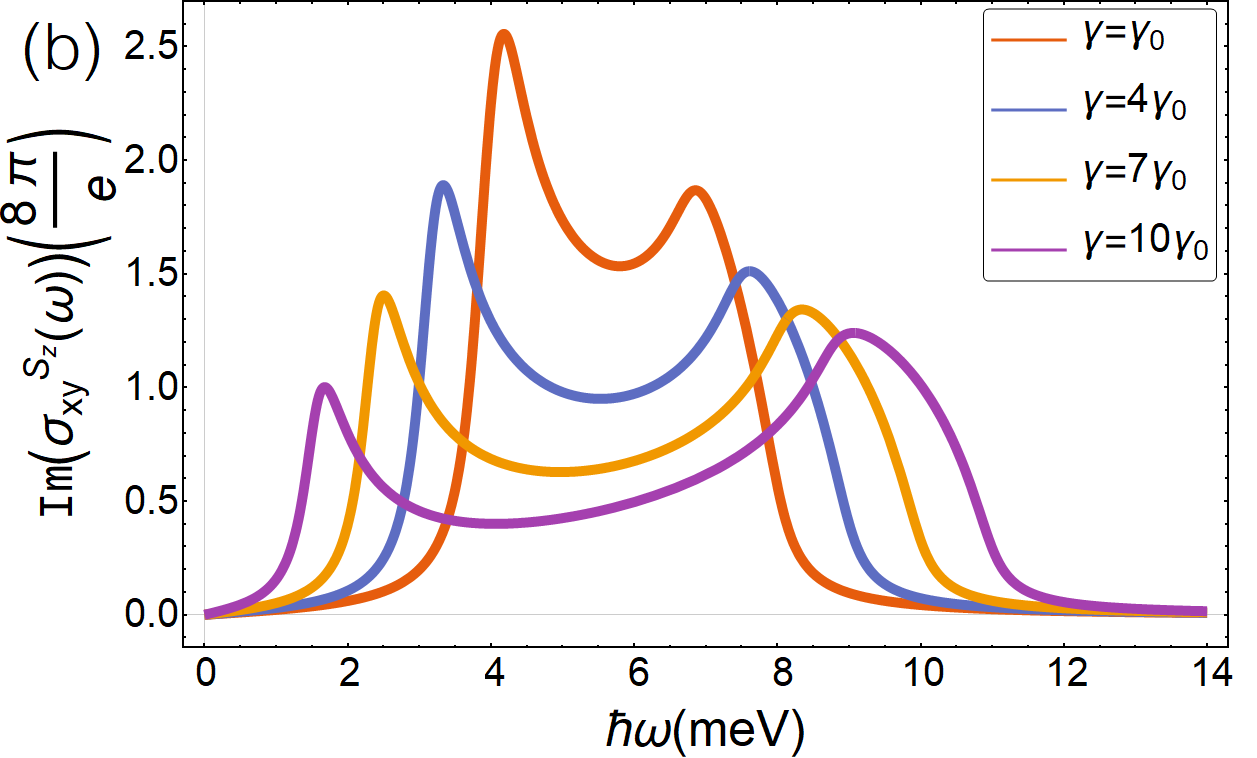}
    \includegraphics[width=7.5cm]{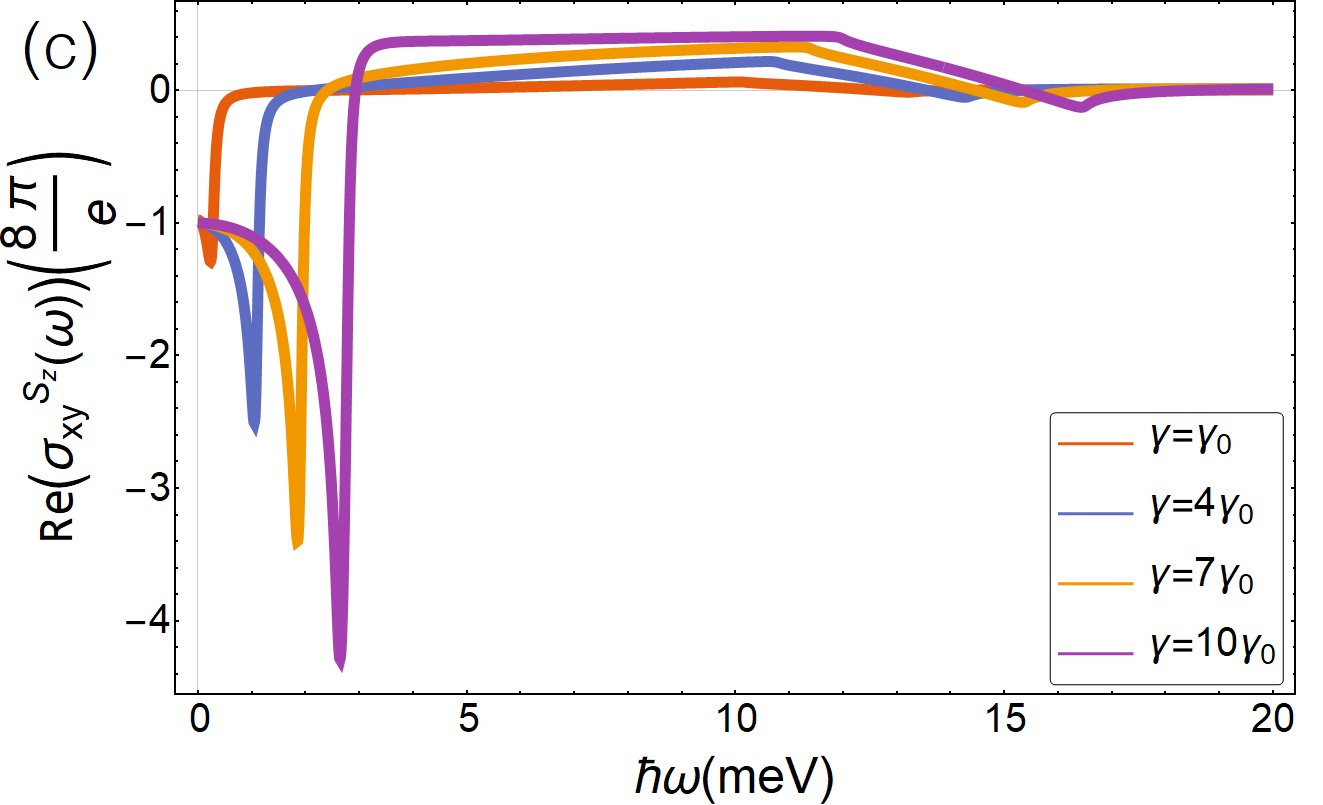}
    \includegraphics[width=7.5cm]{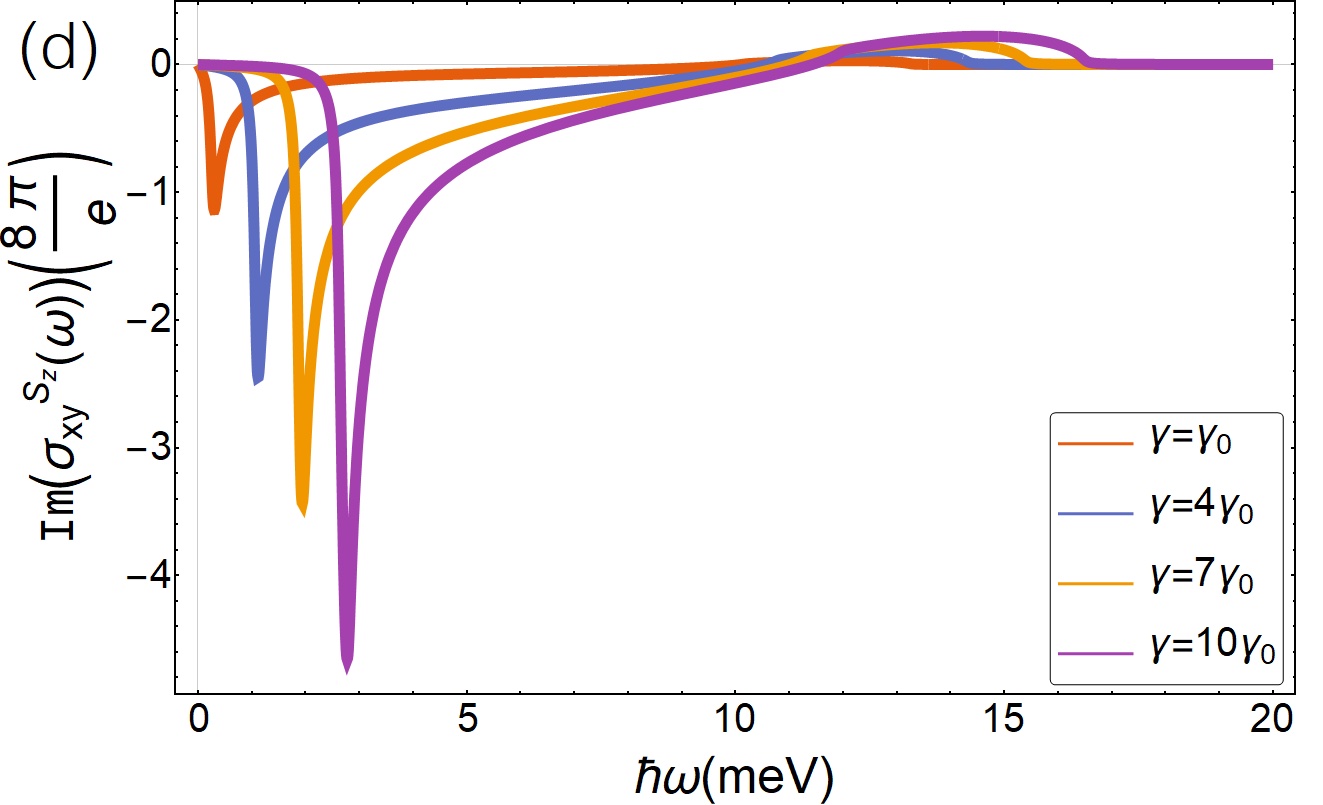}
    \includegraphics[width=7.5cm]{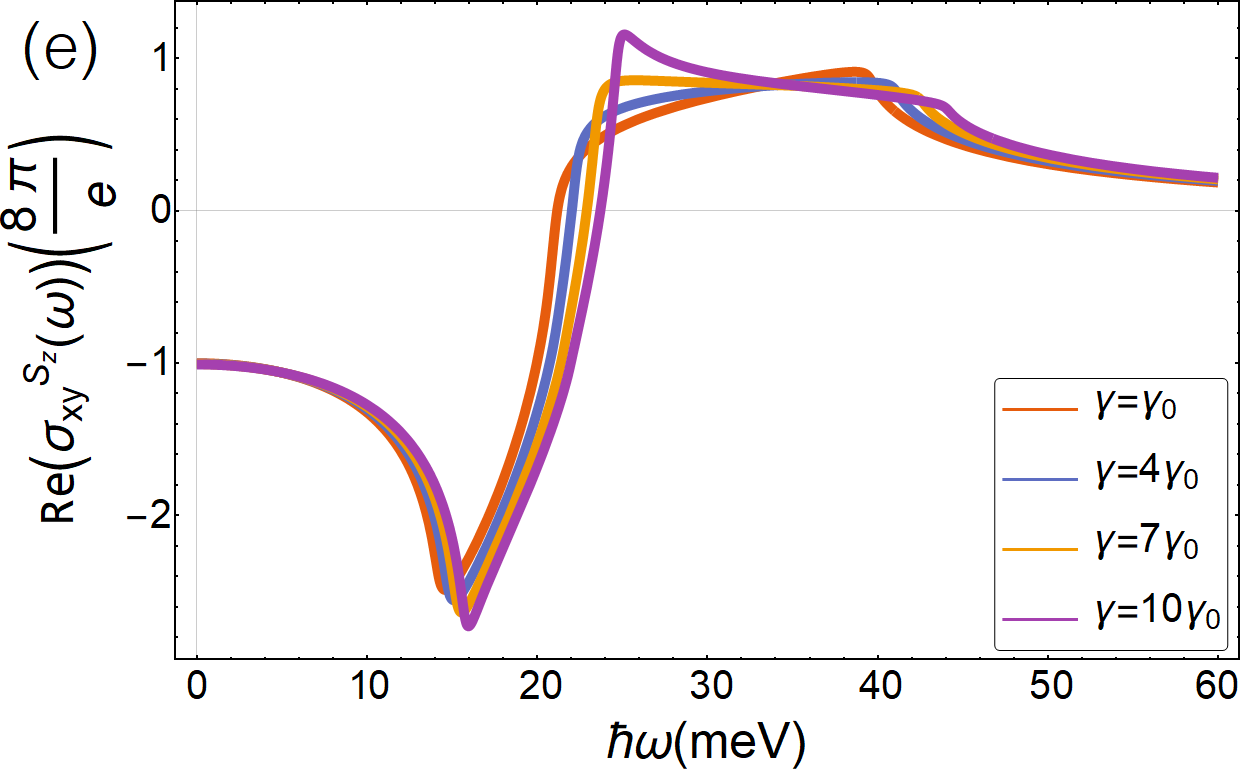}
    \includegraphics[width=7.5cm]{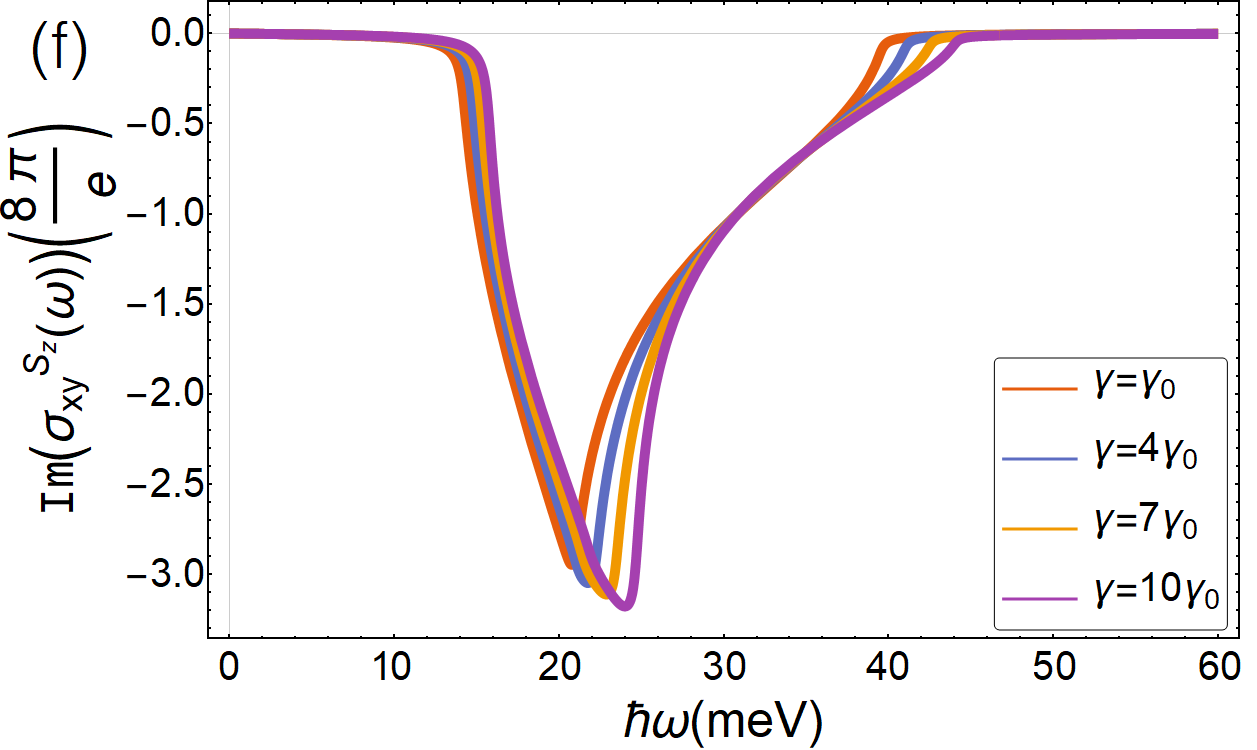}
    \caption{(Colour online) Real and imaginary parts of the spin Hall conductivity for a system with Rashba and Dresselhaus (linear and cubic) SOC. The  spectra were computed for different $\gamma$ values:  $\gamma=\gamma_0$ (red), $\gamma=4\gamma_0$ (blue), $\gamma=7\gamma_0$ (orange) and $\gamma=10\gamma_0$ (purple). We used $\beta=0.25\alpha$, $\hbar\eta=0.25 \times 10^{-3}$~meV~(a)--(b),  
    	$\beta=\alpha$,  $\hbar\eta=0.6\times10^{-4}$~meV~(c)--(d), and  $\beta=4\alpha$, 
    	$\hbar\eta=0.25\times10^{-3}$~meV~(c)--(d). }
    \label{fig:gammagrande}
\end{figure}

In figure~\ref{fig:gammagrande} we show the calculated spectral response for both the real and imaginary part of the spin Hall conductivity $\sigma_{xy}^{s_z}(\omega)$ for three cases, i) weak linear Dresselhaus ($\beta=0.25\alpha$), figure~\ref{fig:gammagrande}~(a)--(b), ii) equal Rashba and linear Dresselhaus coupling strength ($\alpha=\beta$), figure~\ref{fig:gammagrande}~(c)--(d), and iii)~strong linear Dresselhaus interaction ($\beta=4\alpha$), figure~\ref{fig:gammagrande}~(e)--(f). All cases are calculated for different values of the cubic-Dresselhaus coupling strength ($\gamma=\gamma_0,4\gamma_0,7\gamma_0$ and $10\gamma_0,$). Here, we used the values $\alpha=1$.$6\times{10}^{-11}$~eV~m, and $\gamma_0=4.863\times10^{-29}$~eV~m$^{3}$ to produce these plots. The spectra shown for $\beta=0.25\alpha$ (figures~\ref{fig:gammagrande}~a and~\ref{fig:gammagrande}~b) widen as $\gamma$ increases. This follows from the fact that the energy range in which the optical transitions are allowed increases also with $\gamma$. Notoriously, the spin Hall conductivity peaks/minima exhibit higher values for weak cubic-Dresselhaus coupling $\gamma$. In all cases, the real part of the spin Hall conductivity approaches the constant value of conductivity ($e/8\piup$) for the static limit ($\omega\rightarrow0$) and clean limit, while it approaches zero when $\omega$ takes higher values. As for the imaginary part, when $\omega\rightarrow0$ and for high $\omega$ values, the spin Hall conductivity approaches zero too. It develops a pair of peaks near the characteristic absorption energies $\hbar\omega_{a,b}$.  The  energy values for such absorption energies and for the band-edge energies $\hbar\omega_{\pm}$ are shown in table~\ref{025frec}. 

\begin{table}[h]
	\caption{Absorption and band-edge energies,  $\hbar\omega_+$, $\hbar\omega_a$, $\hbar\omega_b$ and $\hbar\omega_-$ for $\beta=0.25\alpha$ and several $\gamma$ values for the plots of figure~\ref{fig:gammagrande}~(a)--(b). Note that the energy difference between the allowed transitions $\hbar\omega_b-\hbar\omega_a$ and  the  difference of the band-edge energies $\hbar\omega_--\hbar\omega_+$  increases as $\gamma$ does.}
\label{025frec}
\vspace{0.5em}
\centering\
\begin{tabular}{|c|c|c|c|c|}
\hline\
Ratio $\gamma$/$\gamma_0$         & $\hbar\omega_+$ {[}meV{]} & $\hbar\omega_a$ {[}meV{]} & $\hbar\omega_b$ {[}meV{]} & $\hbar\omega_-$ {[}meV{]} \\ \hline\
1   & 3.82                      & 4.19                     & 6.70                      & 7.94                      \\ \hline\
4  & 3.15                      & 3.40                     & 7.27                      & 8.75                      \\ \hline\
7  & 2.46                      & 2.61                     & 7.83                      & 9.57                      \\ \hline\
10 & 1.76                      & 1.83                     & 8.38                      & 10.40                     \\ \hline
\end{tabular}
\end{table}

Due to the experimental possibility of tuning the  Rashba coupling parameter $\alpha$ through electrostatic gating, it is interesting to analyse the case when\
the Rashba and linear Dresselhaus coupling strength is set equal ($\beta=\alpha$) and explore its interplay with the cubic-Dresselhaus term on the spin Hall response; this is done in figure~\ref{fig:gammagrande}~(c)--(d). The overall shape of the optical spin Hall conductivity differs significantly with respect to the previous case ($\beta=0.25\alpha$) although it preserves the main characteristics of widening of the spectra. However, in contrast to the weak linear Dresselhaus case, as the cubic-Dresselhaus parameter $\gamma$ is increased, an enhancement of the peaks for both the real and imaginary part of $\sigma_{xy}^{s_z}(\omega)$ is developed, showing the dominant influence of the cubic contribution. Finally, we consider the extreme case in which the modulation of the Rashba parameter $\alpha$ is such that $\beta=4\alpha$, figure~\ref{fig:gammagrande}~(e)--(f). For this scenario, when~$\gamma$ increases, the spectra are shifted slightly to the right at higher energies. The Re[$\sigma_{xy}^{s_z}(\omega)$] of the spectra (figure~\ref{fig:gammagrande}~e) exhibits first a minimum and then a maximum in spin Hall conductivity, with an inflection point in between as the frequency increases. The overall behavior changes the sign with respect to the weak coupling scenario. Similarly as it occurs for the latter case, in the static limit ($\omega\rightarrow0$) and free of disorder limit, the real part of the spin Hall conductivity reaches the constant value of conductivity, but with opposite sign ($-e/8\piup$), and tends to zero for larger frequencies. We have also estimated the characteristic energies associated to the absorption optical transitions and to the band-edge energy transitions allowed. These values are shown in  table~\ref{Table2}.  

\begin{table}[h]
\caption{Absorption and band-edge energies,  $\hbar\omega_{\pm}$, $\hbar\omega_a$, and $\hbar\omega_b$ for $\beta=4\alpha$ and several $\gamma$ values for the plots of figure~\ref{fig:gammagrande}~(e)--(f).}
\label{Table2}
\vspace{0.5em}
\centering\
\begin{tabular}{|c|c|c|c|c|c|}
\hline\
Ratio $\gamma/\gamma_0$ & $\hbar\omega_+$ {[}meV{]} & $\hbar\omega_a$ {[}meV{]} & $\hbar\omega_b$ {[}meV{]} & $\hbar\omega_-$ {[}meV{]} & $\hbar\omega_--\hbar\omega_+$ {[}meV{]} \\ \hline\
1                          & 13.80                     & 20.60                     & 19.17                     & 37.84                     & 24.04                                   \\ \hline\
4                          & 14.13                     & 21.37                     & 19.31                     & 38.55                     & 24.42                                   \\ \hline\
7                          & 14.45                     & 22.15                     & 19.45                     & 39.26                     & 24.81                                   \\ \hline\
10                         & 14.77                     & 22.94                     & 19.59                     & 39.98                     & 25.21                                   \\ \hline
\end{tabular}
\end{table}
 
The difference $\hbar\omega_--\hbar\omega_+$ shows us that the energy range of the allowed optical transitions increases as $\gamma$ is increased. This energy range is larger than the one for the $\beta=0.25\alpha$ case, but the increase in the range for the allowed optical transitions between different $\gamma$ values is less than the case $\beta=4\alpha$.\

\section{Summary and conclusions}
We have investigated the spin Hall conductivity in the frequency domain for 2DEGs under Rashba and (linear plus cubic) Dresselhaus spin-orbit coupling using the Kubo formalism in the linear response. We have shown that a simultaneous presence of Rashba and cubic  Dresselhaus SOC in 2DEGs leads to a strongly anisotropic spin splitting of the bands which in turn gives rise to a characteristic frequency dependence of the spin Hall conductivity. Such characteristic frequencies are connected with the absorption and band-edge energies responsible for the absorption and high-density peaks of the joint density of states. We further analyse the spin Hall conductivity in the static limit in the absence of scattering with impurities or disorder and find that in general it depends on the Fermi wave number and on the SOC parameters, in sharp constrast with the pure Rashba case. The significant widening and energy separation of the peaks as well as the change of sign of the spin Hall conductivity response, as the Rashba parameter is varied in 2DEGs with sizable cubic-Dresselhaus coupling, may offer a signature of the presence and impact of a competing Rashba with linear and cubic Dresselhaus SOC. This could be also of interest for the optical control of spin currents in 2DEGs.
 
\section*{Acknowledgements}
 F. M. thanks professor Bertrand Berche for very exciting scientific collaborations in the past few years. We also acknowledge the support of DGAPA-UNAM through the project PAPIIT No. IN113920. 
 
\appendix 

 \section*{Appendix}
	\renewcommand{\theequation}{A.\arabic{equation}}

In this appendix we describe details of the derivation of the ``spin current-charge current'' correlation function (\ref{correlation_function}) for the most general Hamiltonian (\ref{hamiltoniano_total}) that includes the linear and cubic Dresselhaus interaction in coexistence with the Rashba term. We start by calculating the components of the velocity operators in the Schr\"odinger picture using ${\hat{v}}_j=\frac{1}{\ri\hbar}[{\hat{x}}_j,\hat{H}]  $, with $j=x,y$ leading to
\begin{align}
    {\hat{v}}_y(0)&=\frac{{\hat{p}}_y}{m^*}-\frac{1}{\hbar}\big(\alpha {\hat{\sigma}}_x+\beta{\hat{\sigma}}_y\big)+\frac{\gamma}{\hbar^3}\big(2{\hat{p}}_x{\hat{p}}_y{\hat{\sigma}}_x-{\hat{p}}_x^2{\hat{\sigma}}_y\big),\\
    {\hat{v}}_x(0)&=\frac{{\hat{p}}_x}{m^*}+\frac{1}{\hbar}\big(\alpha{\hat{\sigma}}_y+\beta{\hat{\sigma}}_x\big)+\frac{\gamma}{\hbar^3}\big({\hat{p}}_y^2{\hat{\sigma}}_x-2{\hat{p}}_y{\hat{p}}_x{\hat{\sigma}}_y\big).\
\end{align}
Therefore, the spin-current operator (\ref{densidad_corriente_espin}) of $z$-polarized spins flowing along the $x$-axis reads
\begin{eqnarray}
        {\hat{\mathbf{\cal{J}}}}^{s_z}_x(0)&=&\frac{1}{2}\left\lbrace\frac{\hbar}{2}{\hat{\sigma}}_z,\hspace{0.2cm} {\hat{v}}_x\right\rbrace=\frac{1}{2}\left\lbrace\frac{\hbar}{2}{\hat{\sigma}}_z,\hspace{0.2cm} \left(\frac{{\hat{p}}_x}{{m^*}}+\frac{1}{\hbar}\big(\alpha{\hat{\sigma}}_y+\beta{\hat{\sigma}}_x\big)
        +\frac{\gamma}{\hbar^3}\big({\hat{p}}_y^2{\hat{\sigma}}_x-2{\hat{p}}_y{\hat{p}}_x{\hat{\sigma}}_y\big)\right)
        \right\rbrace \nonumber\\
        &=&\frac{\hbar}{2{m^*}}{\hat{\sigma}}_z {\hat{p}}_x,
\end{eqnarray}
after using the property  $\{\sigma_i,\sigma_j\}=2\delta_{i,j}\mathbb{I}$. In the interaction picture, the time-dependent spin current operator is %
\begin{equation}
    {\hat{\mathbf{\cal{J}}}}^{s_z}_x(t)=\frac{\hbar}{2m^*} {\mathrm{e}}^{\ri{\hat{H}}t/{\hbar}} \, {\hat{\sigma}}_z{\hat{p}}_x \, {\mathrm{e}}^{-\ri{\hat{H}}t/{\hbar}},\
\end{equation}
that yields the to commutator
\begin{eqnarray}
       \left[{\hat{\mathbf{\cal{J}}}}^{s_z}_x(t), \,{\hat{v}}_y(0)\right]&=&\left[\frac{\hbar}{2m^*}{\mathrm{e}}^{\ri{\hat{H}}t/{\hbar}}{\hat{\sigma}}_z{\hat{p}}_x{\mathrm{e}}^{-\ri{\hat{H}}t/{\hbar}}, \, \frac{{\hat{p}}_y}{m^*}-\frac{1}{\hbar}\big(\alpha {\hat{\sigma}}_x+\beta{\hat{\sigma}}_y\big)+\frac{\gamma}{\hbar^3}\big(2{\hat{p}}_x{\hat{p}}_y{\hat{\sigma}}_x-{\hat{p}}_x^2{\hat{\sigma}}_y\big)\right]\nonumber\\
        &=& -      \frac{1}{2m^*}{\mathrm{e}}^{\ri{\hat{H}}t/{\hbar}}{\hat{\sigma}}_z {\hat{p}}_x {\mathrm{e}}^{-\ri{\hat{H}}t/{\hbar}} \big(\alpha{\hat{\sigma}}_x+\beta{\hat{\sigma}}_y\big)\nonumber\\&+&\frac{1}{2{m^*}}\big(\alpha{\hat{\sigma}}_x+\beta{\hat{\sigma}}_y\big){\mathrm{e}}^{\ri{\hat{H}}t/{\hbar}}{\hat{\sigma}}_z {\hat{p}}_x {\mathrm{e}}^{-\ri{\hat{H}}t/{\hbar}}\nonumber\\
        &+&\frac{\gamma m^*}{2\hbar^2}{\mathrm{e}}^{\ri{\hat{H}}t/{\hbar}}{\hat{\sigma}}_z {\hat{p}}_x {\mathrm{e}}^{-\ri{\hat{H}}t/{\hbar}} \big(2{\hat{p}}_x{\hat{p}}_y{\hat{\sigma}}_x-{\hat{p}}_x^2{\hat{\sigma}}_y\big) \nonumber\\
        &-&\frac{\gamma m^*}{2\hbar^2}\big(2{\hat{p}}_x{\hat{p}}_y{\hat{\sigma}}_x-{\hat{p}}_x^2{\hat{\sigma}}_y\big){\mathrm{e}}^{\ri{\hat{H}}t/{\hbar}}{\hat{\sigma}}_z {\hat{p}}_x {\mathrm{e}}^{-\ri{\hat{H}}t/{\hbar}},  \
    \label{commutator}
\end{eqnarray}
and taking the expectation value of (\ref{commutator}) of each term using the eigenvectors (\ref{eigenvectores}), we get,
\begin{eqnarray}
        &&\langle{\mathbf{k}, \nu| ({-1}/{2{m^*}}){\mathrm{e}}^{\ri{\hat{H}}t/{\hbar}}{\hat{\sigma}}_z {\hat{p}}_x {\mathrm{e}}^{-\ri{\hat{H}}t/{\hbar}} \big(\alpha{\hat{\sigma}}_x+\beta{\hat{\sigma}}_y\big)|\mathbf{k}, \nu}\rangle \nonumber\\
        &&=\frac{-1}{2{m^*}}\frac{\hbar k_x^2 \nu \ri}{k\Delta(k,\,\theta)}\big(\alpha^2-\beta^2-\alpha\gamma k_x k_y-\beta\gamma k_y^2\big)\exp{\left[\frac{2\ri k\Delta(k,\,\theta)\nu t}{\hbar}\right]},\
\end{eqnarray}
\begin{eqnarray}
        &&\langle{\mathbf{k}, \nu|({-1}/{2{m^*}})\big(\alpha{\hat{\sigma}}_x+\beta{\hat{\sigma}}_y\big){\mathrm{e}}^{\ri{\hat{H}}t/{\hbar}}{\hat{\sigma}}_z {\hat{p}}_x {\mathrm{e}}^{-\ri{\hat{H}}t/{\hbar}}|\mathbf{k}, \nu}\rangle \nonumber\\
        &&=\frac{-1}{2{m^*}}\frac{\hbar k_x^2 \nu \ri}{k\Delta(k,\,\theta)}\big(\alpha^2-\beta^2-\alpha\gamma k_x k_y-\beta\gamma k_y^2\big)\exp{\left[\frac{-2\ri k\Delta(k,\,\theta)\nu t}{\hbar}\right]},\
\end{eqnarray}
\begin{eqnarray}
        &&\langle{\mathbf{k}, \nu|({\gamma}/{2\hbar^2 {m^*}}){\mathrm{e}}^{\ri{\hat{H}}t/{\hbar}}{\hat{\sigma}}_z {\hat{p}}_x {\mathrm{e}}^{-\ri{\hat{H}}t/{\hbar}} \big(2{\hat{p}}_x{\hat{p}}_y{\hat{\sigma}}_x-{\hat{p}}_x^2{\hat{\sigma}}_y\big)|\mathbf{k}, \nu}\rangle \nonumber\\
        &&=\frac{\gamma\hbar}{2{m^*}}\frac{k_x^2\nu \ri}{k\Delta(k,\,\theta)}\big[\alpha k_x k_y+\beta \big(-2k_y^2+k_x^2\big)-\gamma k_x^2 k_y^2\big]\exp{\left[\frac{2\ri k\Delta(k,\,\theta)\nu t}{\hbar}\right]},\
\end{eqnarray}
\begin{eqnarray}
        &&\langle{\mathbf{k}, \nu|-({\gamma}/{2\hbar^2 {m^*}})\big(2{\hat{p}}_x{\hat{p}}_y{\hat{\sigma}}_x-{\hat{p}}_x^2{\hat{\sigma}}_y\big){\mathrm{e}}^{\ri{\hat{H}}t/{\hbar}}{\hat{\sigma}}_z {\hat{p}}_x {\mathrm{e}}^{-\ri{\hat{H}}t/{\hbar}}|\mathbf{k}, \nu}\rangle \nonumber\\
        &&=\frac{\gamma\hbar}{2{m^*}}\frac{k_x^2\nu \ri}{k\Delta(k,\,\theta)}\big[\alpha k_x k_y+\beta \big(-2k_y^2+k_x^2\big)-\gamma k_x^2 k_y^2\big]\exp{\left[\frac{-2\ri k\Delta(k,\,\theta)\nu t}{\hbar}\right]},
\end{eqnarray}
with $\Delta(k,\theta)$ as defined in (\ref{delta_k_theta}). Hence, finally after regrouping terms, the ``spin current-charge current'' correlation function takes the form (\ref{correlation_function}), \
 \
\
\
\begin{eqnarray}
        \langle{\mathbf{k}, \nu|[{\hat{\mathbf{\cal{J}}}}^{s_z}_x(t), {\hat{v}}_y(0)]|\mathbf{k}, \nu}\rangle&=&\frac{\nu \ri \hbar k_x^2}{k\Delta(k, \theta)m^*}\cos\left[\frac{2k\Delta(k, \theta)t}{\hbar}\right] \nonumber\\
        &\times&\left(-\alpha^2+\beta^2+\alpha\gamma k^2\sin2\theta+\beta\gamma k^2 \cos2\theta-\frac{\gamma^2}{4}k^4 \sin^2 2\theta\right).
\end{eqnarray}
\

\ukrainianpart

\title{Вплив {\it p}-кубічного члена Дрессельгауза на спіновий ефект Холла}
\author{Е. Сантана-Суарес\refaddr{label1},
	Ф. Мірелес\refaddr{label2}}

\addresses{
	\addr{label1}Інститут фізики, Національний автономний університет Мексики (UNAM), PO Box 20-364, 01000 Мехіко, Мексика
	\addr{label2} Фізичний факультет, Центр нанотехнологій, Автономний національний університет Мехіко, 
	22800 Енсенада, Баха Каліфорнія, Мексика}
\makeukrtitle

\begin{abstract}
	\tolerance=3000%
		Добре відомо, що спін-орбітальна взаємодія Дрессельгауза (СОВ) у напівпровідникових двовимірних електронних газах (2D-ЕГ) має як лінійний, так і кубічний внески за імпульсом. Тим не менше, останнім зазвичай нехтують у більшості теоретичних досліджень.	Однак нещодавні експерименти з обертання площини поляризації Керра виявили значне посилення кубічної взаємодії Дрессельгауза шляхом збільшення швидкості дрейфу в 2D-ЕГ, поміщених у квантові ями GaAs. У даній роботі проведено дослідження оп\-тич\-ної спінової провідності Холла у 2D-ЕГ за одночасної присутності СОВ Рашби та Дрессельгауза (за наявності як лінійного, так і кубічного внесків). Дослідження проведені методом Кубо в рамках теорії лінійного відгуку. Показано, що співіснування СОВ Рашби та кубічного внеску СОВ Дрессельгауза у 2D-ЕГ сприяє сильній анізотропії спінового розщеплення зони, що, у свою чергу, призводить до дуже специфічної частотної залежності спінової провідності Холла. Виявлено, що спінова провідність Холла може бути дуже чутливою до значної кубічної складової в інтенсивності СОВ Дрессельгауза. Це може бути актуальним для оптичного контролю спінових струмів у 2D-ЕГ з кубічною СОВ Дрессельхауза.

\keywords{двовимірний електронний газ, спін-орбітальна взаємодія, спіновий ефект Холла, спінова провідність, спіновий перенос}
	
\end{abstract}

\lastpage
\end{document}